\documentclass[11pt]{article}
\usepackage{latexsym}

\usepackage{epsfig}
\usepackage{graphicx}

\catcode`\@=11

\@addtoreset{equation}{section}

\def\@normalsize{\@setsize\normalsize{15pt}\xiipt\@xiipt
\abovedisplayskip 14pt plus3pt minus3pt%
\belowdisplayskip \abovedisplayskip
\abovedisplayshortskip  \z@ plus3pt%
\belowdisplayshortskip  7pt plus3.5pt minus0pt}
\def\small{\@setsize\small{13.6pt}\xipt\@xipt
\abovedisplayskip 13pt plus3pt minus3pt%
\belowdisplayskip \abovedisplayskip
\abovedisplayshortskip  \z@ plus3pt%
\belowdisplayshortskip  7pt plus3.5pt minus0pt
\def\@listi{\parsep 4.5pt plus 2pt minus 1pt
            \itemsep \parsep
            \topsep 9pt plus 3pt minus 3pt}}

\def\underline#1{\relax\ifmmode\@@underline#1\else
        $\@@underline{\hbox{#1}}$\relax\fi}
\@twosidetrue \relax

\catcode`@=12

\catcode`\@=11

\def\section{\@startsection{section}{1}{\z@}{3.5ex plus 1ex minus
   .2ex}{2.3ex plus .2ex}{\large\bf}}

\def\ps@headings{\def\@oddfoot{}\def\@evenfoot{}
\def\@oddhead{\hbox{}\hfill
        \makebox[.5\textwidth]{\raggedright\ignorespaces --\thepage{}--
        \hfill }}
\def\@evenhead{\@oddhead}
\def\subsectionmark##1{\markboth{##1}{}}
}

\ps@headings

\catcode`\@=12

\newcommand{\be}{\begin{equation}}
\newcommand{\ee}{\end{equation}}
\newcommand{\bea}{\begin{eqnarray}}

\newcommand{\eea}{\end{eqnarray}}

\begin{document}

\begin{titlepage}
\begin{flushright}
hep-th/xxxxxxx
\end{flushright}
\begin{centering}
\vspace{.8in}
%


{\large {\bf AdS/CFT Correspondence and the Reheating of the
Brane-Universe}}
\\

\vspace{.5in} {\bf  E. Papantonopoulos$^{1,2*}$ and  V.
Zamarias$^{2\dag}$}

\vspace{0.3in}
$^{1}$Institute of Cosmology and Gravitation, University of
Portsmouth, Portsmouth PO1 2EG, UK
$^{2}$Department of Physics, National Technical University of Athens,\\
Zografou Campus GR 157 73, Athens, Greece\\
\end{centering}


\begin{abstract}

We present a mechanism for exit from inflation and reheating using
the AdS/CFT correspondence. A cosmological evolution is induced on
a probe D3-brane as it moves in a black D-brane background of
type-0 string theory. If the tachyon field is non zero, inflation
is induced on the brane-universe, with the equation of state
parameter in the range $-1<w<-1/3$, depending on the position of
the probe brane in the bulk. As the probe brane approaches the
horizon of the background black hole, the inflation rate decreases
and the value of $w$ gets larger. At some critical distance away
from the horizon, inflation ends. When the brane-universe reaches
the horizon, the conformal invariance is restored, the background
geometry becomes $AdS_{5}\times S^{5}$, and the brane-universe
feels the CFT thermal radiation and reheats.

 \end{abstract}

\vspace{1.0cm}

\begin{flushleft}

  $^{*}$e-mail address: lpapa@central.ntua.gr \\
$^{\dag}$e-mail address: zamarias@central.ntua.gr

\end{flushleft}
\end{titlepage}

\section{Introduction}
All recent observational data supports the idea that our universe
in its early cosmological evolution had an inflationary phase. The
most conventional inflationary dynamics is described by a scalar
field, the inflaton, with a self-interacting potential
\cite{inflation} and the exit from inflation and reheating of the
universe happens when the inflaton field reaches the minimum of
the potential and starts oscillating around that minimum. This
inflationary scenario can cure the problems of the standard
cosmology, like the horizon, flatness and monopole problems, and
the inflationary dynamics parametrized by the inflaton field is
adequate to describe most of the observational data. However, it
is very difficult to produce a complete theoretical model where
all the inflationary dynamics comes out in a consistent way from
the theory \cite{lazarides}. Specifically, understanding the
mechanism of the exit from inflation and reheating is of
fundamental interest in the inflationary scenario.

In this direction, it is important to apply new theoretical ideas
for understanding the inflationary scenario and the mechanisms
involved. Recently, motivated by string theory, the idea of
AdS/CFT correspondence was proposed. The AdS/CFT correspondence is
build on Maldacena conjecture \cite{malda}, that the
(3+1)-dimensional world-volume of N coinciding D3-branes, in the
large N limit, is dual to type IIB superstrings, propagating on
the near-horizon $AdS_{5}\times S^{5}$ background geometry
\cite{gibb}. A very important consequence of this conjecture is
that, because the (3+1)-dimensional world-volume of N coinciding
D3-branes give rise to $\mathcal{ N}=4$ supersymmetric SU(N)
Yang-Mills (SYM) theory \cite{horow,kleb,witten,witten2}, the
thermodynamics of Yang-Mills theory is linked with the
thermodynamics of Schwarzschild black holes embedded in the AdS
space \cite{page}.

Many ideas about AdS/CFT correspondence were influenced by the
intriguing concept of "holography" \cite{hooft,sussk}. The
underlying principle, which was originated in the Bekenstein bound
\cite{bekens}, is based on the notion that the maximal entropy
that can be stored within a given volume will be determined by the
largest black hole fitting inside that volume. Since the entropy
of a black hole is essentially given by its surface area, it
follows directly that all the relevant degrees of freedom of any
system must in some sense live on the boundary enclosing that
system.

Recently, using the holographic principle, the entropy bounds in a
radiation dominated closed Friedmann-Robertson-Walker universe was
analyzed \cite{verlide}. It was found a surprising similarity
between Cardy's entropy formula for 1+1 dimensional CFT and the
Friedmann equation governing the evolution of the universe. After
a suitable identification, it was shown that actually the Cardy's
formula \cite{cardy} maps to the Friedmann equation. In a further
development \cite{savonije} this correspondence between Cardy's
formula and the Friedmann equation was tested in the
Randall-Sundrum type model \cite{randall}. In the case where the
bulk is a Schwarschild-AdS background and there is no matter on
the brane, the correspondence between Cardy's formula and the
Friedmann equation is recovered when the brane crosses the black
hole horizon.

We generalized these results in \cite{zamarias} where, using the
AdS/CFT correspondence we studied  the holographic principle
 in the near-horizon $AdS_{5}\times S^{5}$ geometry with
a probe D3-brane playing the r$\hat{o}$le of the boundary to this
space. The motion of the probe D3-brane in the bulk, induces a
cosmological evolution on the brane. As the brane crosses the
horizon of the bulk Schwarzschild-AdS$_{5}$ black hole, it probes
the holography of the dual CFT. We tested the holographic
principle and we found corrections to the entropy relations in
various physical cases: for radially moving, spinning and
electrically charged D3-brane and for a NS/NS B-field in the bulk.

The cosmological evolution on the brane due to its motion in
higher dimensional spacetimes was studied in \cite{Cha} where the
Israel matching conditions
 were used to relate the bulk to the domain wall (brane) metric, and
 some interesting cosmological solutions were found. In
 \cite{Keh} a universe three-brane is considered in motion
 in ten-dimensional space in the presence of a gravitational field
 of other branes. It was shown that this motion in ambient space
 induces cosmological expansion (or contraction) on our universe,
 simulating various kinds of matter. This is known as Mirage
 Cosmology. Using the technics of Mirage cosmology we showed in \cite{pap1,pap2},
that if a probe D3-brane moves in a non-conformal background of
type-0 string theory, with constant values of dilaton and tachyon
fields, a cosmological evolution is induced on the brane-universe,
which for some range of the parameters has an inflationary phase
and there is also an exit from this phase \cite{pappa1}. In
\cite{Kim}, using similar technics  the
 cosmological evolution of the three-brane in the background of
 type IIB string theory was also considered.

Type-0 string theory is interesting because it does not have
space-time supersymmetry and as a result of GSO projection a
tachyon field appears in the action \cite{tset,Minah,Typ0}.
Actually the tachyon field manifests itself in two ways: through
its tachyon potential and a function f(T) which couples to the RR
flux of the background. It was shown that if this function has an
extremum then, because of its coupling to the RR field, stabilizes
the tachyon potential driving its mass to positive values
\cite{tset,Kleb1}.

Another appealing feature of type-0 string theory is that, when
the tachyon field is non trivial, because of its coupling to
dilaton field, there is a renormalization group flow from infrared
(IR) to ultra violet (UV) \cite{Minah,bigazzi}. This corresponds
to a flow of the couplings of the dual gauge theory from strong
(IR region) to weak couplings (UV region) where the dilaton field
gets small and the tachyon field requires a constant value.
However, because the evolution equations are complicated when the
tachyon and dilaton fields are not constant, the analytic
evolution from IR to UV fixed points is not known, and only the
behaviour of the theory near the fixed points is well understood.
At these two fixed points it has been shown that the background
geometry asymptotes to the near-horizon $AdS_{5}\times S_{5}$
geometry \cite{tset,Typ0}.

In this work we will use the AdS/CFT correspondence to propose a
mechanism for exit from inflation and reheating in the case of a
brane-universe moving in a type-0 string background
\cite{pap1,pap2,pappa1,steer}. We will first show that as the
brane moves in a conformal $AdS_{5}\times S^{5}$ background, there
is no inflation induced on the brane-universe. If the background
is generated by the field of other Dp-branes then inflation is
possible. In particular we will show that when a probe D3-brane is
moving in a black D-brane type-0 string background, inflation is
induced on the brane-universe when the tachyon field is not zero,
and inflation has its maximum rate when it condenses. During
inflation, the equation of state parameter $w$ has the value
$w=-1$ when the brane is away from the horizon of the background
black hole, while it is increasing, as the brane is approaching
the horizon. We will also show, that there is a critical distance
from the horizon where inflation on the brane-universe stops and
the equation of state parameter becomes $w=-1/3$. When the brane
reaches the horizon, the equation of state parameter becomes
$w=1/3$, the background geometry is $AdS_{5}\times S^{5}$ and then
because of the AdS/CFT correspondence, the brane-universe feels
the thermal radiation and reheats.

The paper is organized as follows. After the introduction in Sect.
I, we briefly describe the technics of Mirage cosmology in Sect.
II and we apply these technics to a conformal near-horizon
$AdS_{5}\times S^{5}$ background. In Sect. III we discuss the case
of a probe D3-brane moving in a general background generated by
other Dp-branes and in particular the brane motion in the D-brane
type-0 string background. In Sect. IV we discuss the inflation
generated on the brane-universe and the mechanism to exit from it
and reheating, and finally in Sect. V are our conclusions.

\section{A Probe D3-Brane Moving in a Near-Horizon $\mathbf{AdS_{5}\times S^{5}}$
Black Hole Background}

Consider a probe D3-brane moving in a generic
 static, spherically symmetric background.
 We assume the brane to be light compared to the background so
 that we will neglect the back-reaction.
The background metric we consider has the general form
\begin{equation}\label{in.met}
ds^{2}_{10}=g_{00}(r)dt^{2}+g(r)(d\vec{x})^{2}+
  g_{rr}(r)dr^{2}+g_{S}(r)d\Omega_{5},
\end{equation}
 where $g_{00}$ is negative, and there is also a dilaton field $\Phi$ as well as a $RR$
 background~$C(r)=C_{0...3}(r)$ with a self-dual field strength.

The dynamics on the brane will be governed by the
 Dirac-Born-Infeld action  given by
\begin{eqnarray}\label{B.I. action}
  S&=&T_{3}~\int~d^{4}\xi
  e^{-\Phi}\sqrt{-det(\hat{G}_{\alpha\beta}+(2\pi\alpha')F_{\alpha\beta}-
  B_{\alpha\beta})}
  \\  \nonumber
  &+&T_{3}~\int~d^{4}\xi\hat{C}_{4}+anomaly~terms.
\end{eqnarray}
 The induced metric on the brane is
\begin{equation}\label{ind.metric}
  \hat{G}_{\alpha\beta}=G_{\mu\nu}\frac{\partial x^{\mu}\partial x^{\nu}}
  {\partial\xi^{\alpha}\partial\xi^{\beta}},
\end{equation}
 with similar expressions for $F_{\alpha\beta}$ and
 $B_{\alpha\beta}$.
For an observer on the brane the Dirac-Born-Infeld action is the
volume of the brane trajectory modified by the presence of the
anti-symmetric two-form $ B_{\alpha\beta}$, and worldvolume
anti-symmetric gauge fields $ F_{\alpha\beta}$.
 As the brane moves
 the induced world-volume metric becomes a function of
 time, so there is a cosmological evolution from the brane point
 of view \cite{Keh}.

  In the static
 gauge, $x^{\alpha}=\xi^{\alpha},\alpha=0,1,2,3 $
 using (\ref{ind.metric}) we can calculate the bosonic part of the
 brane Lagrangian which reads
\begin{equation}\label{brane Lagr}
\mathcal{L}=\sqrt{A(r)-B(r)\dot{r}^{2}-D(r)h_{ij}\dot{\varphi}^{i}\dot{\varphi}^{j}}
-C(r),
\end{equation}
where $h_{ij}d \varphi ^{i} d \varphi^{j}$ is the line
 element of the unit five-sphere, and
\begin{equation}\label{met.fun}
  A(r)=g^{3}(r)|g_{00}(r)|e^{-2\Phi},
  B(r)=g^{3}(r)g_{rr}(r)e^{-2\Phi},
  D(r)=g^{3}(r)g_{S}(r)e^{-2\Phi},
\end{equation}
and $C(r)$ is the $RR$ background. The problem is effectively
one-dimensional and can be solved easily. The momenta are given by
\begin{eqnarray}
p_{r}&=&-\frac{B(r)\dot{r}}{\sqrt{A(r)-B(r)\dot{r}^{2}}},\nonumber
\\
p_{i}&=&-\frac{D(r)h_{ij}\dot{\phi}^{j}}{{\sqrt{A(r)-B(r)\dot{r}^{2}-
D(r)h_{ij}\dot{\phi}^{i}\dot{\phi}^{j}}}}. \end{eqnarray} Since
(\ref{brane Lagr}) is not explicitly time dependent and the
$\phi$-dependence is confined to the kinetic term for
$\dot{\phi}$, for an observer in the bulk, the brane moves in a
geodesic parametrised by a conserved energy $E$ and a conserved
angular momentum $l^{2}$ given by
\begin{eqnarray}
E&=&\frac{\partial\mathcal{L}}{\partial \dot{r}}\dot{r}+
\frac{\partial\mathcal{L}}{\partial
\dot{\phi}^{i}}\dot{\phi}^{i}-\mathcal{L}=p_{r}\dot{r}+p_{i}\dot{\phi}^{i}-\mathcal{L},\nonumber
\\ l^{2}&=&h^{ij}\frac{\partial\mathcal{L}}{\partial
\dot{\phi}^{i}} \frac{\partial\mathcal{L}}{\partial
\dot{\phi}^{j}}=h^{ij}p_{i}p_{j}.
\end{eqnarray}
Solving these expressions for $\dot{r}$ and $ \dot{\phi}$ we find
\begin{equation}\label{functions}
\dot{r}^{2}=\frac{A}{B}(1-\frac{A}{(C+E)^{2}}\frac{D+\ell^{2}}{D}),
\,\,
h_{ij}\dot{\varphi}^{i}\dot{\varphi}^{j}=\frac{A^{2}\ell^{2}}{D^{2}(C+E)^{2}}.
\end{equation}
The allowed values of $r$ impose the constraint that $C(r)+E\geq
0$. The induced four-dimensional metric
 on the brane, using (\ref{ind.metric}) in the static gauge, is
\begin{equation}\label{fmet}
d\hat{s}^{2}=(g_{00}+g_{rr}\dot{r}^{2}+g_{S}h_{ij}\dot{\varphi}^{i}\dot{\varphi}^{j})dt^{2}
+g(d\vec{x})^{2}.
\end{equation}
In the above relation we substitute $ \dot{r}^{2}$ and $
h_{ij}\dot{\varphi}^{i}\dot{\varphi}^{j} $ from (\ref{functions}),
and using (\ref{met.fun}) we get
\begin{equation}
d\hat{s}^{2}=-\frac{g_{00}^{2}g^{3}e^{-2\phi}}{(C+E)^{2}}dt^{2}+g(d\vec{x})^{2}.
\end{equation} We can define the cosmic time $\eta$  as
\begin{equation}\label{cosmic}
 d\eta=\frac{|g_{00}|g^{\frac{3}{2}}e^{-\Phi}}{|C+E|}dt,
\end{equation} so the induced metric becomes
\begin{equation}\label{fin.ind.metric}
d\hat{s}^{2}=-d\eta^{2}+g(r(\eta))(d\vec{x})^{2},
\end{equation}

 The induced metric on the brane (\ref{fin.ind.metric}) is the standard form of a flat
expanding universe. We can derive the analogue of the
four-dimensional Friedmann equations by defining $g=\alpha^{2}$

\begin{equation}\label{dens}  \Big{(}\frac
{\dot{\alpha}}{\alpha}\Big{)}^{2}=
\frac{(C+E)^{2}g_{S}e^{2\Phi}-|g_{00}|(g_{S}g^{3}+\ell^{2}e^{2\Phi})}
{4|g_{00}|g_{rr}g_{S}g^{3}}\Big{(}\frac{g'}{g}\Big{)}^{2},
\end{equation}
 where the dot stands for derivative with respect to cosmic time
 while the prime stands for derivatives with respect to $r$. The
 right hand side of (\ref{dens}) can be interpreted in terms of an
 effective matter density on the probe brane
 \begin{equation}\label{denseff1} \frac{8\pi G}{3}\rho_{eff}=
\frac{(C+E)^{2}g_{S}e^{2\Phi}-|g_{00}|(g_{S}g^{3}+\ell^{2}e^{2\Phi})}
{4|g_{00}|g_{rr}g_{S}g^{3}}\Big{(}\frac{g'}{g}\Big{)}^{2},
\end{equation} where $G$ is the four-dimensional Newton's
constant. We can also calculate \begin{eqnarray}  \label{dadot}
\frac{\ddot{\alpha}}{\alpha}&=&\Big{(}1+\frac{g}{g'}\frac{\partial}{\partial
r}\Big{)}
\frac{(C+E)^{2}g_{S}e^{2\Phi}-|g_{00}|(g_{S}g^{3}+\ell^{2}e^{2\Phi})}
{4|g_{00}|g_{rr}g_{S}g^{3}}\Big{(}\frac{g'}{g}\Big{)}^{2}\\
\nonumber &=&\Big{[}1+\frac{1}{2}\alpha\frac{\partial} {\partial
\alpha}\Big{]}\frac{8\pi G}{3}\rho_{eff}.
\end{eqnarray}
If we set the above equal to $-\frac{4\pi G
}{3}(\rho_{eff}+3p_{eff})$ we can define the effective pressure
$p_{eff}$.

 Therefore, the motion of a D3-brane on a
general spherically symmetric background had induced on the brane
an energy density and a pressure. Then, the first and second
Friedmann equations can be derived giving a cosmological evolution
of the brane-universe in the sense that an observer on the brane
measures a scale factor $\alpha(\eta)$ of the brane-universe
evolution. This scale factor depends on the position of the brane
in the bulk. This cosmological evolution is known as "Mirage
Cosmology" \cite{Keh}: the cosmological evolution is not due to
energy density on our universe but on the energy content of the
bulk.

We will apply the above described formalism to the near-horizon
conformal geometry $AdS_{5}\times S^{5}$. There are
Schwarzschild-AdS$_{5}$ black hole solutions in this background
 with metric
\begin{equation} ds^{2}=\frac{r^{2}}{L^{2}}\Big{(}-f(r)dt^{2}+(d
\vec{x}) ^{2}\Big{)}
+\frac{L^{2}}{r^{2}}\frac{dr^{2}}{f(r)}+L^{2}d \Omega ^{2}_{5},
\label{bhmetric} \end{equation} where $f(r)=1-\Big{(}
\frac{r_{0}}{r}\Big{)}^{4} $. The $RR$ field is given by
$C=C_{0...3}= \Big{[}
\frac{r^{4}}{L^{4}}-\frac{r^{4}_{0}}{2L^{4}}\Big{]}$.

Using (\ref{in.met}) we find in this background
\begin{eqnarray} g_{00}(r)&=&-\frac{r^{2}}{L^{2}} \Big{(}1-\big{(}
\frac{r_{0}}{r} \Big{)}^{4} \Big{)} =-\frac{1}{g_{rr}} \nonumber
\\ g(r)&=&\frac{r^{2}}{L^{2}} \nonumber \\ g_{s}(r)&=& L^{2},
\label{bachground}
\end{eqnarray} and the brane-universe scale factor is $ \alpha=r/L $.
Substituting the above functions to equation (\ref{denseff1}) we
find the analogue of Friedmann equation on the brane which is
\cite{Keh}
\begin{equation}
H^{2}=\frac{8\pi G}{3}\rho_{eff}=\frac{1}{L^{2}} \Big{[}
\Big{(}1+\frac{1}{\alpha^{4}}\Big{(}E-r_{0}^{4}/2L^{4}\Big{)}
\Big{)}^{2}-\Big{(}1-\Big{(}\frac{r_{0}}{L}\Big{)}^{4}
\frac{1}{\alpha^{4}}\Big{)}
\Big{(}1+\frac{l^{2}}{L^{2}}\frac{1}{\alpha^{6}}\Big{)}
 \Big{]}, \label{friedbh} \end{equation} where
E is a constant of integration of the background field equations,
expressing the conservation of energy, and it is related to the
black hole mass of the background \cite{steer}, while
$r_{0}^{4}/2L^{4}$ is the constant part of the RR field,
expressing essentially electrostatic energy, and it can be
absorbed into the energy $
 \tilde{E}=E-r_{0}^{4}/2L^{4}$. This Friedmann equation
 describes the cosmological evolution of a contracting or
 expanding universe depending on the motion of the probe brane. This
 motion in turn depends on two parameters the energy $\tilde{E}$
 and the angular momentum $l^{2}$. These two parameters specify various
 trajectories of the probe brane. The scale factor $\alpha$ comes
 in various powers, indicating that (\ref{friedbh}) describes the cosmological
 evolution of various
 kind of Mirage or stiff cosmological matter.

 Defining the dimensionless parameter $
a=l^{2}/{L^{2}}\alpha^{6} $, equation (\ref{friedbh}) becomes
\begin{equation} H^{2}=\frac{8\pi G}{3}\rho_{eff}=\frac{1}{L^{2}}
\Big{[} \Big{(}1+\frac{\tilde{E}}{\alpha^{4}}
\Big{)}^{2}-\Big{(}1-\Big{(}\frac{r_{0}}{L}\Big{)}^{4}
\frac{1}{\alpha^{4}}\Big{)} \Big{(}1+a \Big{)}
 \Big{]}. \label{friedbha} \end{equation}

Using equations (\ref{dadot}) and (\ref{friedbha}), the second
Friedmann equation in this background reads
\begin{equation} \label{friedgsec}
\dot{H}=-\frac{2}{L^{2}}
\Big{[}2\frac{\tilde{E}}{\alpha^{4}}\Big{(}
1+\frac{\tilde{E}}{\alpha^{4}}
\Big{)}+\Big{(}\frac{\alpha_{0}}{\alpha}
\Big{)}^{4}\Big{(}1+a\Big{)}-\frac{3}{2}a\Big{(}1-\Big{(}\frac{\alpha_{0}}{\alpha}
\Big{)}^{4}\Big{)} \Big{]},
\end{equation} and the effective pressure, using again (\ref{dadot}), is
\begin{equation}
p_{eff}=\frac{1}{8 \pi GL^{2}}
\Big{[}\Big{(}\frac{\alpha_{0}}{\alpha}
\Big{)}^{4}+5\frac{\tilde{E}^{2}}{\alpha^{8}}
+2\frac{\tilde{E}}{\alpha^{4}} +7a\Big{(}\frac{\alpha_{0}}{\alpha}
\Big{)}^{4}-3a    \Big{]}. \label{effpres}
\end{equation} From (\ref{friedbha}) we also have the
effective energy density
\begin{equation}
\rho_{eff}=\frac{3}{8 \pi GL^{2}}
\Big{[}\Big{(}\frac{\alpha_{0}}{\alpha}
\Big{)}^{4}+\frac{\tilde{E}^{2}}{\alpha^{8}}
+2\frac{\tilde{E}}{\alpha^{4}}-a\Big{(}
1-\Big{(}\frac{\alpha_{0}}{\alpha} \Big{)}^{4}\Big{)} \Big{]}.
\label{effnerg}
\end{equation}
It is instructive to consider the equation of state
$p_{eff}=w\rho_{eff}$ where $w$ is given by
\begin{equation}
w=\frac{1}{3}\Big{[}\frac{\Big{(}\frac{\alpha_{0}}{\alpha}
\Big{)}^{4}+5\frac{\tilde{E}^{2}}{\alpha^{8}}
+2\frac{\tilde{E}}{\alpha^{4}} +7a\Big{(}\frac{\alpha_{0}}{\alpha}
\Big{)}^{4}-3a }{\Big{(}\frac{\alpha_{0}}{\alpha}
\Big{)}^{4}+\frac{\tilde{E}^{2}}{\alpha^{8}}
+2\frac{\tilde{E}}{\alpha^{4}}-a\Big{(}
1-\Big{(}\frac{\alpha_{0}}{\alpha} \Big{)}^{4}\Big{)}}\Big{]}.
\label{ww}
\end{equation}
As the brane moves in the Schwarzschild-AdS$_{5}$  black hole
background, the equation of state is parametrized by the energy of
the bulk and the angular momentum of the brane. However, when
$\tilde{E}=0$ and $a=0$ the brane-universe is radiation dominated
at any position in the bulk \cite{zamarias}, as can be seen from
(\ref{ww}). This is expected, because the only scale in the theory
is the energy scale $\title{E}$ and putting it to zero the theory
is scale invariant, while a non-zero angular momentum induces on
the brane all kind of Mirage matter. This can also be understood
with the use of the AdS/CFT correspondence. The moving probe
D3-brane is playing the r$\hat{o}$le of the boundary of the
$AdS_{5}\times S^{5}$ geometry. Because the background space is an
AdS space, the dual theory on the brane is CFT, therefore the
brane at any position would be in a thermal radiation state.

We can also investigate if an inflationary phase can be generated
on the brane-universe as it moves in the  near-horizon
$AdS_{5}\times S^{5}$ geometry. Defining the acceleration
parameter $I$ as \be I=H^{2}+\dot{H} \label{accel} \ee and using
(\ref{friedbha}) and (\ref{friedgsec}), and putting $a=0$, we get
\be I=-\frac{1}{L^{2}}\Big{[}3\frac{\tilde{E}}{\alpha^{4}} +
2\frac{\tilde{E}^{2}}{\alpha^{8}}
+\frac{\alpha_{0}^{4}}{\alpha^{4}}\Big{]}. \ee Therefore for
positive energy and no angular momentum on the brane, there is no
inflation induced on the brane-universe. Observe that for
$\tilde{E}=0$, $I$ is always negative irrespective of the brane
position. To have inflation on the brane the background should not
be conformal as we will discuss in the following.

\section{The Non-Conformal Type-0 String Background}

In this section we will consider more general non-conformal
backgrounds. Such backgrounds are described by Dp-black branes
with a metric \cite{horow}
\begin{equation}
ds^{2}_{10}=\frac{1}{\sqrt{H_{p}}}\Big{(}-f(r)dt^{2}+(d \vec{x})
^{2}\Big{)} +\sqrt{H_{p}}\,\frac{dr^{2}}{f(r)}+ \sqrt{H_{p}}\,
r^{2}d \Omega ^{2}_{8-p}, \label{bhmetricdp}
\end{equation}
where $ H_{p}=1+\Big{(} \frac{L}{r}\Big{)} ^{7-p}$, \, and
$f(r)=1-\Big{(} \frac{r_{0}}{r}\Big{)} ^{7-p}$. In this background
the RR form is
\begin{equation}
C_{012...p}=\sqrt{1+\Big{(} \frac{r_{0}}{L}\Big{)} ^{7-p}}\,
\frac{1-H_{p}(r)}{H_{p}(r)},
\end{equation} and the dilaton field takes the form $ e^{\Phi}=H_{p}^
{(3-p)/4}$. Taking the near horizon limit of the above geometry,
we recover the Schwarzschild-Ad$S_{5}\times S^{5}$ black hole
geometry discussed in the previous section. The technics of Mirage
cosmology can be applied to these general backgrounds \cite{Keh}
and the Friedmann equation on the brane can be derived. In
particular, we will discuss the D-brane background of type-0
string theory.

The action of the type-0 string is given by \cite{tset}
\begin{eqnarray}\label{action}
S_{10}&=&~\int~d^{10}x\sqrt{-g}\Big{[} e^{-2\Phi} \Big{(}
 R+4(\partial_{\mu}\Phi)^{2} -\frac{1}{4}(\partial_{\mu}T)^{2}
-\frac{1}{4}m^{2}T^{2}-\frac{1}{12}H_{\mu\nu\rho}H^{\mu\nu\rho}\Big{)}
\nonumber \\&& - \frac{1}{4}(1+T+\frac{T^{2}}{2})|F_{5}|^{2}
\Big{]}.
\end{eqnarray}
The tachyon is coupled to the $RR$ field through the function
\begin{equation}\label{ftac}
 f(T)=1+T+\frac{1}{2} T^{2}.
\end{equation}

The tachyon field appearing in (\ref{action}) is a result of GSO
projection and there is no spacetime supersymmetry in the theory.
The tachyon field appears in (\ref{action}) through its kinetic
term, its potential and through the tachyon function $f(T)$. The
potential term is giving a negative mass squared term which
however can be shifted to positive values if the function $f(T)$
has an extremum. This happens in the background where the tachyon
field acquires vacuum expectation value $T_{vac}=-1$ and
$f^{\prime}(T_{vac})=0$ \cite{tset,Kleb1}. In this background the
dilaton equation is \be \nabla^{2}\Phi=-\frac{1}{4\alpha^{\prime}}
e^{\frac{1}{2}\Phi}T^{2}_{vac}. \ee This equation is giving a
running of the dilaton field which means that the conformal
invariance is lost, and $AdS_{5}\times S^{5}$ is not a solution.
Therefore, the tachyon condensation is responsible for breaking
the 4-D conformal invariance of the theory and as we will see in
the next section it also induces inflation on a probe-brane moving
in this background. However, the conformal invariance is restored
in two conformal points, corresponding to IR and UV fixed points,
when the tachyon field gets a constant value. The flow from IR to
UV as exact solutions of the equations of motion derived from
action (\ref{action}) is not known, but when the dilaton and
tachyon fields are constant such a solution can be found
\cite{tset,Typ0}
 which can be extended to a black D-brane \cite{adi}
\begin{equation}
ds^{2}_{10}=\frac{1}{\sqrt{H}}\Big{(}-\phi(r)dt^{2}+(d \vec{x})
^{2}\Big{)} +\sqrt{H}\,\frac{dr^{2}}{\phi(r)}+ \sqrt{H}\, r^{2}d
\Omega ^{2}_{5}, \label{bhmetrictyp0}
\end{equation}
where $ H=1+\Big{(} \frac{e^{\Phi_{0}}Q}{2r^{4}}\Big{)}$,  and
$\phi(r)=1-\Big{(} \frac{r_{0}}{r}\Big{)} ^{4}$. Q is the electric
RR charge and $ \Phi_{0}$ denotes a constant value of the dilaton
field. If we define $L=\Big{(}e^{\Phi_{0}}Q/2\Big{)}^{1/4}$ we can
write $ H=1+\Big{(} \frac{L}{r}\Big{)}^{4}$.

In this non-conformal background inflation can be induced on the
brane-universe. The case of a constant dilaton and tachyon field
was studied in \cite{pap1}, while in \cite{pap2} approximate
solutions of the equations of motion were used to study the
induced on the brane inflationary phase.

\section{AdS/CFT Correspondence, Exit from Inflation and Reheating}

In this section we will study the way the inflation ends and
reheating starts on the brane-universe using the AdS/CFT
correspondence. In the background metric (\ref{bhmetrictyp0})
using (\ref{in.met}) we find
\bea g_{00}&=&-H^{-1/2}\phi(r) \nonumber \\
g&=&H^{-1/2}=\alpha^{2}=\Big{(}1+\Big{(}\frac{L}{r}\Big{)}^{4}\Big{)}^{-1/2}\nonumber \\
g_{rr}&=&
H^{1/2}\phi^{-1}(r)\nonumber \\
g_{s} &=&H^{1/2}r^{2} \label{gg1} \eea and the RR field is given
by
$C=e^{-\Phi_{0}}f^{-1}(T)\Big{(}1+\Big{(}\frac{L}{r}\Big{)}^{4}\Big{)}^{-1}+Q_{1}$
where $Q_{1}$ is an integration constant. Substituting (\ref{gg1})
into (\ref{denseff1}) we get the induced Friedmann equation on the
brane \be
 H^{2}=\frac{(1-\alpha^{4})^{5/2}}{L^{2}}
\Big{[} \Big{(}f^{-1}(T)+\frac{\tilde{E}}{\alpha^{4}}
\Big{)}^{2}-\Big{(}1-\Big{(}\frac{\alpha_{0}}{\alpha}\Big{)}^{4}
\Big{(}\frac{1-\alpha^{4}}{1-\alpha^{4}_{0}}\Big{)}\Big{)}
\Big{(}1+a\Big{(}1-\alpha^{4}\Big{)}^{1/2}e^{2\Phi_{0}} \Big{)}
 \Big{]}, \label{friedbha1} \end{equation} where
 $\tilde{E}=(Q_{1}+E)e^{\Phi_{0}}$. To simplify the discussion we
 consider the case where the constant value of the dilaton field
 is zero and also the angular momentum of the probe brane is
 also zero, $a=0$. In this case from (\ref{friedbha1}) we can calculate
 the second Friedmann equation and then the equation of state
\bea w=\frac{p_{eff}}{\rho_{eff}}&=&\frac{10}{3}\alpha^{4}
 (1-\alpha^{4})^{-1}+\frac{1}{3}\Big{\{}\Big{[}8\frac{\tilde{E}}{\alpha^{4}}\Big{(}f^{-1}(T)+\frac{\tilde{E}}{\alpha^{4}}
\Big{)}-3\Big{(}f^{-1}(T)+\frac{\tilde{E}}{\alpha^{4}} \Big{)}^{2}
 \nonumber \\
&+&3\Big{(}1-\Big{(}\frac{\alpha_{0}}{\alpha}\Big{)}^{4}
\Big{(}\frac{1-\alpha^{4}}{1-\alpha^{4}_{0}}\Big{)}\Big{)}+4\Big{(}\frac{\alpha_{0}}{\alpha}
\Big{)}^{4}\Big{(}1-\alpha^{4}_{0} \Big{)}^{-1} \Big{]}\nonumber \\
&\Big{/}&\Big{[}\Big{(}f^{-1}(T)+\frac{\tilde{E}}{\alpha^{4}}
\Big{)}^{2}-\Big{(}1-\Big{(}\frac{\alpha_{0}}{\alpha}\Big{)}^{4}
\Big{(}\frac{1-\alpha^{4}}{1-\alpha^{4}_{0}}\Big{)}\Big{)}
\Big{]}\Big{\}}. \label{eqstate} \eea This equation depends on the
energy parameter $\tilde{E}$. We know that if we take the
near-horizon limit of (\ref{bhmetrictyp0}) the geometry becomes
exact $AdS_{5}\times S^{5}$ and then because of the AdS/CFT
correspondence the dual theory on the brane must be CFT. We use
this fact to fix the free parameter $\tilde{E}$. Considering that
the bulk black hole is near extremal and taking the near-horizon
limit $\alpha<<1$ of (\ref{eqstate}) we find \be
w=\frac{1}{3}\Big{(}\frac{5\frac{\tilde{E}^{2}}{\alpha^{8}}
+2\frac{\tilde{E}}{\alpha^{4}}f^{-1}(T)+3(1-f^{-2}(T))+\frac{\alpha^{4}_{0}}{\alpha^{4}}}
{f^{-2}(T)+2f^{-1}(T)\frac{\tilde{E}}{\alpha^{4}}+\frac{\tilde{E}^{2}}{\alpha^{8}}+\frac{\alpha^{4}_{0}}{\alpha^{4}}
-1} \Big{)}. \label{w1} \ee Demanding to have $w=1/3$ we get \be
\frac{\tilde{E}}{\alpha^{4}}=\pm\Big{(}f^{-2}(T)-1 \Big{)}^{1/2}.
\label{energycond} \ee Note that if the tachyon field is zero,
also $\tilde{E}=0$, then from (\ref{w1}), $w=1/3$. This is because
with $T=0$ we recover the near-horizon $AdS_{5}\times S^{5}$
geometry and then the results of Sect. II are valid. We know that
we do not have inflation induced on the brane in the near-horizon
$AdS_{5}\times S^{5}$ geometry, therefore we parametrize the
departure from the near-horizon limit by \be
\frac{\tilde{E}}{\alpha^{4}}=\pm\Big{(}f^{-2}(T)-1 \Big{)}^{1/2}
\frac{\alpha^{4}_{0}}{\alpha^{4}}~. \label{denergycond} \ee In the
above relation we have used the characteristic scale in the bulk
which is set by the horizon of the black hole. The departure from
the near-horizon limit is then equivalent to how far away the
probe brane is from the horizon of the bulk black hole. We
substitute (\ref{denergycond}) into (\ref{friedbha1}) (for
$\Phi_{0}=a=0$) and we get \be
H^{2}=\frac{1}{L^{2}}\Big{[}f^{-2}(T)\Big{(}1+
\Big{(}\frac{\alpha_{0}}{\alpha} \Big{)}^{8} \Big{)}\pm
2\Big{(}\frac{\alpha_{0}}{\alpha}
\Big{)}^{4}f^{-1}(T)\Big{(}f^{-2}(T)-1 \Big{)}^{1/2}-1
-\Big{(}\frac{\alpha_{0}}{\alpha} \Big{)}^{8}
+\Big{(}\frac{\alpha_{0}}{\alpha} \Big{)}^{4}\Big{]}.
\label{offlimh} \ee Using the above Friedmann equation we can
calculate the effective energy density and pressure on the brane
and then the equation of state becomes $p_{eff}=w \rho_{eff}$,
where \bea w=\frac{1}{3}\Big{\{}\frac{f^{-2}(T)\Big{(}-3+
5\Big{(}\frac{\alpha_{0}}{\alpha} \Big{)}^{8} \Big{)}\pm
2\Big{(}\frac{\alpha_{0}}{\alpha}
\Big{)}^{4}f^{-1}(T)\Big{(}f^{-2}(T)-1 \Big{)}^{1/2}+3
-5\Big{(}\frac{\alpha_{0}}{\alpha} \Big{)}^{8}
+\Big{(}\frac{\alpha_{0}}{\alpha} \Big{)}^{4}}
 {f^{-2}(T)\Big{(}1+
\Big{(}\frac{\alpha_{0}}{\alpha} \Big{)}^{8} \Big{)}\pm
2\Big{(}\frac{\alpha_{0}}{\alpha}
\Big{)}^{4}f^{-1}(T)\Big{(}f^{-2}(T)-1 \Big{)}^{1/2}-1
-\Big{(}\frac{\alpha_{0}}{\alpha} \Big{)}^{8}
+\Big{(}\frac{\alpha_{0}}{\alpha} \Big{)}^{4} }   \Big{\}}.
\label{dw} \eea Observe that on the horizon $\alpha_{0}=\alpha$,
 we get $w=1/3$, independently of the value of the tachyon field, as expected.
The acceleration parameter (\ref{accel}), for positive energy,
using (\ref{offlimh}) is \be
I=\frac{1}{L^{2}}\Big{[}f^{-2}(T)\Big{(}1-3
\Big{(}\frac{\alpha_{0}}{\alpha} \Big{)}^{8} \Big{)}-
2\Big{(}\frac{\alpha_{0}}{\alpha}
\Big{)}^{4}f^{-1}(T)\Big{(}f^{-2}(T)-1 \Big{)}^{1/2}-1
+3\Big{(}\frac{\alpha_{0}}{\alpha} \Big{)}^{8}
-\Big{(}\frac{\alpha_{0}}{\alpha} \Big{)}^{4}\Big{]}.
\label{accelofflimh} \ee
\begin{figure}[t]
\centering
\hspace{0.1cm}%
\includegraphics[scale=1.0]{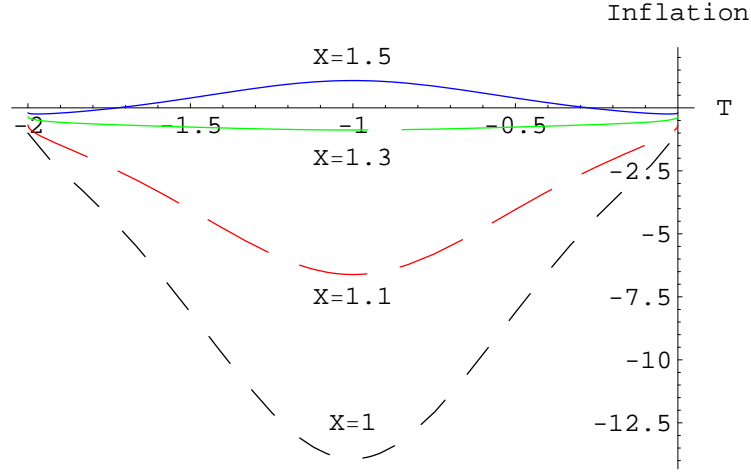}
 \caption{Inflation as a function of the tachyon field T,  in various distances from the horizon,
  where X$=\alpha/\alpha_{0}$.}
\end{figure}

In Fig. 1 we plotted the acceleration parameter I as a function of
the tachyon field. Inflation can start at a critical distance
$\alpha\sim1.37\alpha_{0}$ from the horizon and depends on the
value of the tachyon field. During inflation the tachyon field
gives the maximum rate of inflation when it condenses at
$T_{vac}=-1$. In Fig. 2 we plotted the acceleration parameter as a
function of equation of state $w$ parameter.  At the critical
value $\alpha=1.37\alpha_{0}$, when tachyon condenses, $w=-1/3$
and inflation starts. For distances $\alpha > 1.37\alpha_{0}$ $w$
is decreasing and for larger values of the distance goes
asymptotically to $w=-1$. For $\alpha\ < 1.37\alpha_{0}$ where
there is no inflation, $w$ is increasing and on the horizon takes
the value $w=1/3$.

\begin{figure}[t]
\centering
\hspace{0.1cm}%
\includegraphics[scale=1.0]{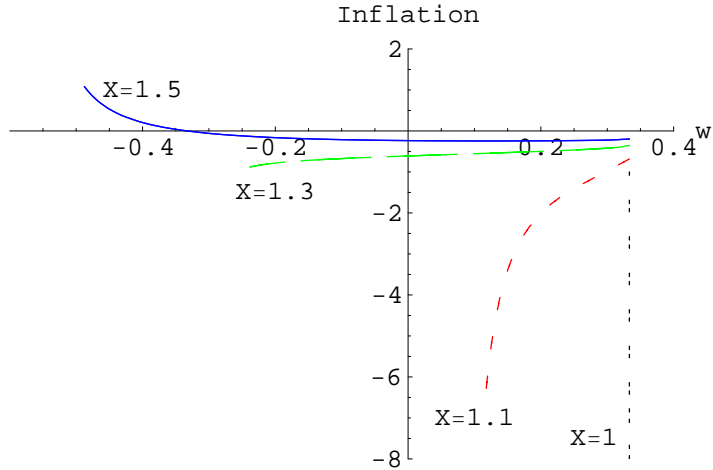}
 \caption{Inflation as a function of the equation of state parameter w.
  Again here X$=\alpha/\alpha_{0}$.}
\end{figure}


The physical picture we get from our previous analysis is as
follows. We can trust the theory only near the conformal points
where the $AdS_{5}\times S^{5}$ geometry is restored. We
parametrized the departure from the exact $AdS_{5}\times S^{5}$
geometry by the ratio $\alpha/\alpha_{0}$ where $\alpha_{0}$ is
the horizon of the bulk black hole. As the inflating probe brane
approaches the black hole horizon the inflation rate decreases and
the equation of state parameter $w$ increases departing from its
asymptotic value $w=-1$. At some critical distance from the
horizon, depending on the value of the tachyon field, the probe
brane stops inflating and at that moment $w=-1/3$. As the probe
brane comes nearer to the horizon, $w$ increases, and on the
horizon, where the near-horizon $AdS_{5}\times S^{5}$ geometry is
restored, $w=1/3$ and the brane-universe reheats.

\section{Conclusions}

We have presented a mechanism for exit from inflation and
reheating using the AdS/CFT correspondence. We followed the motion
of a probe D3-brane moving in a black type-0 string background.
For constant values of the dilaton and tachyon fields, inflation
is induced on the brane-universe. At relative large distances from
the horizon of the background black hole, the equation of state
parameter $w$ has its minimum value $w=-1$, while inflation has
its maximum rate when the tachyon field condenses.  As the brane
comes nearer to the horizon, the inflation rate decreases and the
value of $w$ gets larger. At some critical distance away from the
horizon, inflation ends. When the brane-universe reaches the
horizon, the conformal invariance is restored, the background
geometry becomes $AdS_{5}\times S^{5}$, the dual theory on the
brane is CFT and the brane-universe reheats.

The presence of the closed tachyon field in the background, is
crucial to the generation of inflation on the brane-universe. If
$T=0$ the theory has an exact conformal invariance, the background
geometry is $AdS_{5}\times S^{5}$ and there is no inflation
induced on the brane-universe. The tachyon field couples to the RR
flux of the background with the function $f(T)$. Because of this
coupling, when the tachyon field condenses, the theory is
stabilized. It also couples to the dilaton field and because of
that, the tachyon condensation is responsible for breaking the 4-D
conformal invariance of the theory. When the tachyon field
condenses, the induced inflation on the brane-universe has its
maximum rate. However, the tachyon function $f(T)$ is arbitrary
and it is not predicted by the theory. In this work we considered
a polynomial form of the function, but it would be interesting to
consider other forms, like exponential, and study its effect to
the cosmological evolution on the brane-universe.

We have considered the simplest case of a probe brane moving
radially in the D-brane type-0 string background. A non zero
angular momentum on the probe brane is another source of breaking
the conformal invariance of the theory and it would be interesting
to see what is the effect of a non zero angular momentum on the
brane inflation. Another possible extension of our work is to
consider an electric field on the brane. The electric field on the
brane acts effectively as radiation term on the brane-universe and
it is interesting to see its effect on the reheating.

\section{Acknowlegements}

We acknowledge a very helpful correspondence with A. Armoni. E.P
would like to thank the Institute of Cosmology and Gravitation, of
the University of Portsmouth, for its hospitality, where this work
was completed. Work partially supported by NTUA research program
"Protagoras" and by the Greek Ministry of Education program
"Hraklitos".

\end{document}